# A versatile numerical approach for calculating the fracture toughness and R-curves of cellular materials

Meng-Ting Hsieh[1], Vikram S. Deshpande[3], Lorenzo Valdevit[1,2] [*]

[1]*Mechanical and Aerospace Engineering Dept., University of California, Irvine, CA 92697, USA*

[2]*Materials Science and Engineering Dept., University of California, Irvine, CA 92697, USA*

[3]*Department of Engineering, University of Cambridge, Cambridge CB2 1PZ, UK*

**Abstract**

We develop a numerical methodology for the calculation of mode-I R-curves of brittle and elastoplastic lattice materials, and unveil the impact of lattice topology, relative density and constituent material behavior on the toughening response of 2D isotropic lattices. The approach is based on finite element calculations of the J-integral on a single-edge-notch-bend (SENB) specimen, with individual bars modeled as beams having a linear elastic or a power-law elasto-plastic constitutive behavior and a maximum strain-based damage model. Results for three 2D isotropic lattice topologies (triangular, hexagonal and kagome) and two constituent materials (representative of a brittle ceramic (silicon carbide) and a strain hardening elasto-plastic metal (titanium alloy)) are presented. We extract initial fracture toughness and R-curves for all lattices and show that (i) elastic brittle triangular lattices exhibit toughening (rising R-curve), and (ii) elasto-plastic triangular lattices display significant toughening, while elasto-plastic hexagonal lattices fail in a brittle manner. We show that the difference in such failure behavior can be explained by the size of the plastic zone that grows upon crack propagation, and conclude that the nature of crack propagation in lattices (brittle vs ductile) depends both on the constituent material and the lattice architecture. While results are presented for 2D truss-lattices, the proposed approach can be easily applied to 3D truss and shell-lattices, as long as the crack tip lies within the empty space of a unit cell.

---

[*] Corresponding Author. E-mail: Valdevit@uci.edu

## 1. Introduction

The mechanical properties of cellular materials (i.e., architected materials consisting of a solid and a void phase) depend on the nature of the constituent material and the topology of the architecture. The optimal design of the architecture to achieve maximum stiffness and strength has been extensively investigated over the past two decades, both numerically and experimentally. While a number of efficient designs have been identified, both properties are limited by theoretical bounds: for any cellular material, the relative Young's modulus, $\bar{E} = E/E_s$ and the relative yield strength, $\bar{\sigma}_y = \sigma_y/\sigma_{ys}$ (with the subscript *s* denoting the properties of the constituent materials) cannot exceed the relative density of the material, $\bar{\rho} = \rho/\rho_s$ (Voigt bound). In the case of isotropic cellular materials, much tighter bounds exist, namely the Hashin-Shtrikman bound (Hashin and Shtrikman, 1963) and the Suquet-Ponte-Castaneda nonlinear bound (Castaneda and Debotton, 1992; Suquet, 1993), respectively. A number of nearly-isotropic topologies that achieve or approach the bounds have been identified, e.g., the plate-based closed cell architected material with cube-octet unit cell (Berger et al., 2017) and stochastic shell-based architected materials with spinodal topology (Hsieh et al., 2019).

Conversely, the fracture toughness of cellular materials is theoretically unbounded, and has been much less investigated. The first systematic investigation dates back to 1984, when Maiti, Gibson, and Ashby derived the analytical expressions of mode I fracture toughness, $K_{Ic}$, for both open and closed-cell brittle isotropic foams, by using dimensional analysis that relates the global stress intensity factor, $K$, to the local microscopic stress in the cell wall (Maiti et al., 1984); fracture occurs ($K = K_c$) when the maximum stress in the cell wall around the crack tip reaches the tensile strength, $\sigma_{TS}$, of the constituent material. $K_{IC}$ of open and closed-cell isotropic foams is found to scale with $\bar{\rho}^{1.5}$ and $\bar{\rho}^2$, respectively (Maiti et al., 1984). Later, Gibson and Ashby showed that the $K_{IC}$ of 2D elastic brittle hexagonal lattices scales with $\bar{\rho}^2$ (Gibson and Ashby, 1988). Huang and Lin used the same approach to obtain analytical expressions for mode II fracture toughness, $K_{IIC}$, and mixed mode fracture toughness for both 2D hexagonal lattices and 3D elastic brittle isotropic foams (Huang and Lin, 1996). These models predict that for all cellular materials, the initial fracture toughness can be expressed as $K_{IC} = D\,\sigma_{TS}\,\sqrt{\ell}\,\bar{\rho}^d$, with $\ell$, the unit cell size, and $d$ and $D$, non-dimensional parameters that depend on the topology. While the exponent $d$ can be derived from analytical considerations, the pre-factor $D$ is in general extracted by numerical simulations or experiments.

Alternatively, one can evaluate the fracture toughness analytically using a enriched continuum Cosserat (micropolar) theory; the continuum model is developed by equating its strain energy to that of discrete lattices, in analogy with strain gradient theories (Fleck et al., 1994; Fleck and Hutchinson, 1996, 1993). The first applications of the enriched continuum theory to lattice structures were proposed by Banks and Sokolowski (Banks and Sokolowski, 1968) and Bazant and Christensen (Bazant and Christensen, 2011) . Using this method, Chen *et al*. obtained the expressions of **$K_{IC}$** for 2D triangular, square, and hexagonal lattices, predicting the exponent value

$d = 1$ (Chen et al., 1998). Such prediction for hexagonal lattices is in disagreement with $d = 2$ obtained previously by dimensional analysis and experiment data (Gibson and Ashby, 1988); this discrepancy is attributed to the assumption of affine deformation (stretching) implicit in the Cosserat medium calculations, which ignores cell wall bending (Fleck and Qiu, 2007). As an additional limitation, enriched continuum techniques, as all techniques utilizing homogenization, lack the ability to capture local instabilities in the individual struts of discrete lattices, such as those induced by elastic buckling (Quintana-Alonso and Fleck, 2009).

A number of numerical techniques, based on the Finite Elements method, have been introduced to calculate the fracture toughness of cellular materials (including the pre-factor *D*). Discrete lattices are modeled in configurations with known relations of stress intensity factor to remote boundary conditions such as single edge notched bend (SENB) or center-cracked plate specimens, and treated as a framework of rigid-jointed interconnected beams; the stresses in individual beams as a function of the remotely applied loads are then calculated using the finite element method and the fracture toughness is obtained under the same linear elastic fracture mechanics (LEFM) assumption implicit in the analytical derivations of Gibson and Ashby (Gibson and Ashby, 1988) and Maiti *et al.* (Maiti et al., 1984). Under these assumptions, accurate estimation of fracture toughness requires that the global K-field be sufficiently larger than the local microscopic cell size. The implication is that in the case of cellular materials, the characteristic size of the overall sample, *L*, must be much larger than the crack length, *a*, which in turn must be much larger than the unit cell size, $\ell$. Huang and Gibson investigated 2D elastic brittle diamond lattices and showed that $a/\ell$ must be larger than 7 (Huang and Gibson, 1991). Quintana and Fleck demonstrated the existence of a transitional crack size, $a_T$, such that below it no K-field exists (strength-controlled failure) and above it failure is well predicted by LEFM (toughness-controlled failure). It was shown that

$a_T = 0.9(\ell^2/t)$ for single edge notched bend (SENB) specimens and $a_T = 0.14(\ell^3/t^2)$ for a center-cracked plate subjected to uniaxial tension, with *t* the in-plane thickness of a bar (Quintana-alonso et al., 2010; Quintana-Alonso and Fleck, 2007). The implication is that for $\bar{\rho} < 16\%$, smaller SENB specimens suffice as a fracture toughness testing method. These $a_T$ values also explained the contradictory results between center-cracked plate specimens and SENB specimen of the same 2D lattice found by Huang and Chiang (Huang and Chiang, 1996).

As an alternative to center-cracked plate and SENB specimens, the fracture toughness of 2D lattices can also be obtained via boundary layer analysis. In the simplest term, boundary layer analysis relates the local maximum stress around the macroscopic crack tip to the global asymptotic K-field of a crack (Kanninen and Popelar, 1985; Williams, 1957) in an equivalent homogeneous medium with effective elastic properties, and K-corresponding displacements and rotations are prescribed onto the outer periphery of the finite element mesh. This technique was first applied by Schmidt and Fleck (Schmidt and Fleck, 2001) to gain insights of fracture toughness of regular and irregular elasto-plastic hexagonal lattices. However, extensive usage of such technique is on the study of elastic brittle materials. For example, Fleck and Qiu (Fleck and Qiu, 2007) applied this technique to extract the fracture behaviors of 2D isotropic brittle elastic kagome, triangular, and hexagonal lattices, and demonstrated that $K_{IC}$ scales as $\bar{\rho}^{0.5}$ for kagome lattices, as $\bar{\rho}^1$ for triangular lattices, and as $\bar{\rho}^2$ for hexagonal lattices. The superior fracture behavior of kagome lattices was attributed to a reduced stress level around the crack tip due to elastic crack tip blunting (Fleck et al., 2010). Subsequently, Romijn and Fleck applied the same technique with orthotropic K-field boundary conditions to show that $K_{IC}$ of both square and diamond lattices scale as $\bar{\rho}^1$ (Romijn and Fleck, 2007; Sih et al., 1965). Finally, Christodoulou *et al.* used a similar technique to investigate the effect of cell regularity on the fracture toughness of 2D hexagonal lattices (Christodoulou and Tan, 2013); they found that mode I fracture toughness is more sensitive to topological variations than mode II in the vicinity of the crack tip.

The boundary layer analysis can also be used to investigate the crack initiation and crack propagation of elastoplastic 2D lattices (Schmidt and Fleck, 2001). Under the conditions of small yielding, Schmidt and Fleck applied the displacement boundary conditions corresponding to a $K_I$ field to hexagonal lattices and assumed that the constituent material follows a bilinear hardening law until one of the beams near the crack tip attains the fracture strength, $\sigma_f$. The beam is then gradually removed while the magnitude of the applied K field is increased during the process, such that the work of fracture at the failed joint equals the prescribed fracture energy of the beam. Such requirement of adjusting the applied K field so as to match the prescribed value of fracture energy in a beam disallows the simulation of simultaneous failures of multiple beams. The initial fracture toughness of hexagonal lattices was shown to scale as $\bar{\rho}^2$ and the subsequent R-curves were extracted in both regular and irregular topologies (Schmidt and Fleck, 2001). More recently, Tankasala, Deshpande and Fleck (Tankasala et al., 2015) extracted the initial fracture toughness of elastoplastic 2D lattices (triangular, kagome, diamond, and hexagonal) by combining the boundary layer analysis with two sets of failure criteria: (i) maximum local tensile strain and (ii) average tensile strain. They showed that the predicted fracture toughness is only sensitive to the choice of failure criterion for hexagonal lattices (which are bending-dominated) but not for triangular and kagome lattices (which are stretching-dominated). The initial fracture toughness of elasto-plastic 2D lattices was shown to scale with the relative density with the same power laws as for elastic brittle materials (Tankasala et al., 2015).

We emphasize that all these boundary layer numerical studies make three fundamental assumptions: (i) regardless of the constituent material response (linearly elastic brittle or elasto-plastic), the asymptotic stress field is governed by LEFM; (ii) the local fracture event is predicted using maximum stress/maximum strain criterion governed by a constituent material parameter such as tensile strength $\sigma_{TS}$, fracture strength $\sigma_f$, or fracture strain $\varepsilon_f$; (iii) the crack tip must lie within the empty space of a unit cell. Under assumption (i), the boundary layer technique in ductile materials is limited to small displacement analyses such that the plastic zone size remains confined at the crack tip (small-scale yielding). Furthermore, complexity in applying the K-dependent displacement boundary conditions in the periphery of a finite element mesh could rise significantly in complicated topologies such as those with anisotropic properties. Finally, despite few recent attempts (Choi and Sankar, 2005; Romijn and Fleck, 2007), this technique has yet to be proven as a viable approach for 3D lattice materials.

## 2. The numerical model

### *2.1 Synopsis*

In this work, we combine the maximum strain failure criterion with the single edge notched bend (SENB) specimen configuration to extract the initial fracture toughness, $K_{IC}$, and the R-curve of cellular materials. We apply this technique to three isotropic 2D lattice topologies (triangular, kagome, and hexagonal), two constituent materials (silicon carbide (elastic brittle) and titanium alloy (power-law elasto-plastic)), and three different relative densities $\bar{\rho} = 13\%, 16\%, \text{and } 20\%)$. We verify that our predictions for initial fracture toughness agree with previous works summarized in Table 1 (Fleck, 2009; Tankasala et al., 2015), validating our numerical approach. In addition, we show that scaling relationships for fracture toughness, $K_{IC}$, with relative density only depend on the lattice topology and not on the constituent material (i.e., the same scaling exits for elastic brittle and elastoplastic constituent materials). Furthermore, we demonstrate that triangular lattices exhibit toughening (rising R-curve) even when made of brittle constituent materials. When made of ductile elasto-plastic materials, triangular lattices have the largest initial fracture toughness and the most pronounced R-curves, with multi-stage stable crack propagations and significant spread of plasticity, while hexagonal lattices show nearly no toughening and very low resistance to crack growth. We conclude that the nature of crack propagation in lattices (brittle vs ductile) depends both on the constituent material and the lattice architecture. Finally, we emphasize that the numerical approach presented in this work is applicable to both 2D and 3D architected materials (whether truss or shell-based) made by both brittle and ductile constituent materials, with the only proviso that the crack tip must lie within the empty space of a unit cell (i.e., no stress singularity at the crack tip). These features make this approach ideally suited for further investigations on the toughness characteristics of a wide range of mechanical metamaterials.

| | $D$ for mode-I initial fracture toughness | | | | | Scaling |
|---|---|---|---|---|---|---|
| | Elastic Brittle $D = \frac{K_{IC}}{\sigma_{TS}\sqrt{\ell}\,\bar{\rho}^d}$ | | Elastoplastic $D = \frac{K_{JIC}}{\sigma_{ys}\sqrt{\ell}\,\bar{\rho}^d}(\frac{\varepsilon_{ys}}{\varepsilon_f})^{\frac{P+1}{2P}}$ | | | |
| Topology | I | * | $p = 10^{II}$ | $p = 16^*$ | $p = \infty^{II}$ | $d^{I,II,*}$ |
| Triangular | 0.5 | 0.53 | 0.42 | 0.57 | 0.38 | 1 |
| Kagome | 0.21 | 0.26 | 0.16 | 0.22 | 0.09 | 0.5 |
| Hexagonal | 0.8 | 0.82 | 0.5 | 0.47 | 0.22 | 2 |

*Table 1. Pre-factors, D, and scaling exponents, d, for mode-I initial fracture toughness, $K_{IC}$ (elastic brittle base materials) and $K_{JIC}$ (elastoplastic base materials) of triangular, kagome, and hexagonal lattices. Results are taken from [I] Fleck and Qiu, 2007, [II] Tankasala et al., 2015, *current work. The strain hardening exponent, $p = 16^*$ is obtained by fitting the titanium alloy material properties used in this work to Ramberg-Osgood description. $\ell$ is the bar length. $\varepsilon_{ys}$ is the yield strength, and $\varepsilon_f$ is the fracture strain.*

### 2.2 Specimen design

Proper design of SENB specimens for fracture toughness investigations in cellular materials is dictated by several conditions: (i) the specimen aspect ratio must follow the ASTM E1820 standard for bulk materials (ASTM E1820, 2011), namely $L = 4.5W = 9B$, with $L$ the in-plane specimen length, $W$ the in-plane specimen width, and B the out-of-plane specimen thickness (B is only important if a physical experiment is conducted in accordance with the ASTM standard or in simulations of 3D lattices, but irrelevant in the 2D plane strain simulations discussed in section 2.4); (ii) the number of unit cells must be sufficient to ensure that the K-field exists and is accurately captured (Choi and Sankar, 2005; O'Masta et al., 2017); (iii) the number of elements per bar (in the case of strut-based structures) must be sufficient to allow nonlinear finite strain analysis under large deformations (Tankasala et al., 2015); (iv) the degree of orthotropy of the cellular material must be carefully considered, given that $K$ depends on a dimensionless function of orthotropy (Bao et al., 1992; Quintana-alonso et al., 2010).

In this study, SENB specimens made of three isotropic 2D lattices (kagome, triangular, and hexagonal) were treated as rigid frames of connected bars, ignoring the nodal geometry. The unit cell of each topology is shown in Fig. 1 (a). As these lattices are isotropic, $K$ is only a function of the number of unit cells for a given set of $\ell$, $\bar{\rho}$ and $\sigma_{ys}$. First, we investigated the convergence of mode-I K-field with increasing number of unit cells, as detailed in Appendix A. Second, we investigated the convergence of the load-line displacement at which the maximum axial strain reached the fracture strain $\varepsilon_f$, with increasing number of Timoshenko beam elements per bar, as detailed in Appendix B. The specimens were built by tessellating the unit cells of each topology according to the results from Appendix A. A notch with $a = W/2$ was subsequently cut in the out-of-plane direction by removing bars. We ensured that $a > \ell^2/t$ all densities of all topological specimen, so that failures are toughness-controlled (Quintana-alonso et al., 2010). The final designs of specimens are shown in Fig. 1 (b). See Table 2 for full specimens' specifications tabulated in Table 2.

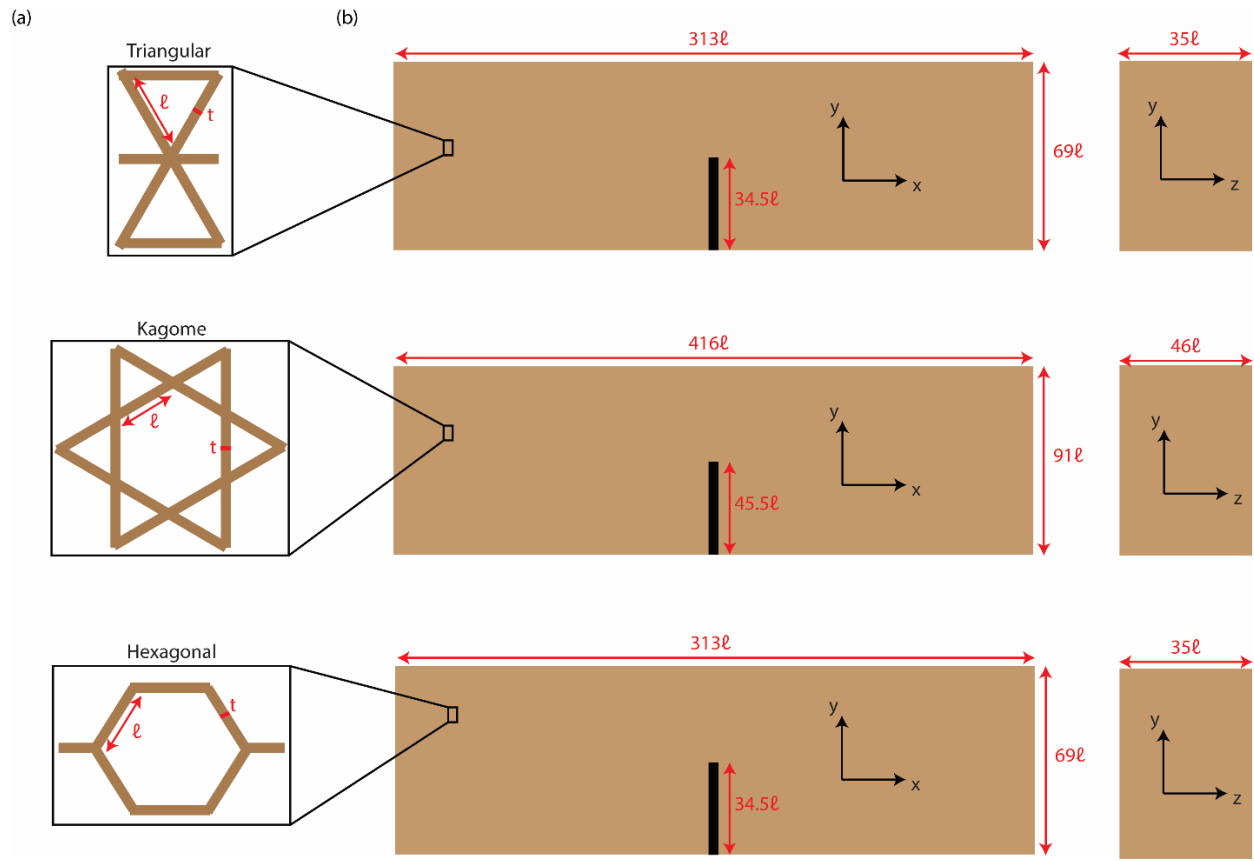

*Fig. 1. (a) Unit cell topology and (b) SENB (single edge notched bend) specimen dimensions of triangular, kagome, and hexagonal specimens from top to bottom. Out-plane thicknesses are shown to illustrate the physical dimensions that would be required to perform a SENB experiment in accordance with ASTM standard; these thicknesses are irrelevant for the 2D plane-strain simulation performed in this work.*

(a)

| ℓ = 3 mm | Triangular | | | Kagome | | | Hexagonal | | |
|---|---|---|---|---|---|---|---|---|---|
| $\bar{p}$ (%) | 13 | 16 | 20 | 13 | 16 | 20 | 13 | 16 | 20 |
| t (mm) | 0.11 | 0.14 | 0.17 | 0.23 | 0.28 | 0.35 | 0.34 | 0.42 | 0.52 |
| a (mm) | | 104 | | | 136 | | | 107 | |
| L (mm) | | 939 | | | 1247 | | | 936 | |
| W (mm) | | 208 | | | 273 | | | 207 | |
| B (mm) | | 105 | | | 139 | | | 104 | |
| $n_y$ | | 40 | | | 45 | | | 40 | |
| $n_e$ | | 4 | | | 20 | | | 30 | |

(b)

| ℓ = 10 mm | Triangular | | | Kagome | | | Hexagonal | | |
|---|---|---|---|---|---|---|---|---|---|
| $\bar{p}$ (%) | 13 | 16 | 20 | 13 | 16 | 20 | 13 | 16 | 20 |
| t (mm) | 0.37 | 0.47 | 0.57 | 0.77 | 0.93 | 1.17 | 1.13 | 1.4 | 1.73 |
| a (mm) | | 347 | | | 453 | | | 357 | |
| L (mm) | | 3130 | | | 4157 | | | 3120 | |
| W (mm) | | 694 | | | 910 | | | 690 | |
| B (mm) | | 350 | | | 464 | | | 347 | |
| $n_y$ | | 40 | | | 45 | | | 40 | |
| $n_e$ | | 4 | | | 20 | | | 30 | |

*Table 2. Specifications of triangular, kagome, and hexagonal specimens with bar length of (a) ℓ = 3 mm and (b) ℓ = 10 mm.*

### *2.3 Constitutive and damage models*

The mechanical behavior of the cell wall materials is described by two models: (i) a constitutive model, which governs the stress-strain behavior of the base material, and (ii) a damage model, which governs fracture of the base material, i.e. material degradation/removal of elements in the finite element analyses, as shown in Fig. 2 (a). The constitutive model is divided in two regions: (a) the linear elastic region, defined by the base material Young's modulus, $E_s$ and Poisson's ratio, $v_s$, and (b) the power-law strain-hardening plastic region, i.e., $\sigma = \sigma_{ys}(1 + \varepsilon_p{}^n)$, with $\sigma_{ys}$ the initial yield strength, $n$ the strain hardening power, and $\varepsilon_p$ the plastic strain of the base material. The damage model is described in terms of the fracture strain, $\varepsilon_f$, of the base material: fracture occurs when the maximum axial strain around the crack tip reaches $\varepsilon_f$. To bracket realistic material behavior, two extremely different base materials were considered: (i) titanium alloy (Ti-6Al-4V), representative of ductile metals, with $E_s$ = 123 GPa, $v_s$ = 0.3, $\sigma_{ys}$ = 932 MPa, $n$ =0.7237 and $\varepsilon_f$ = 0.1105 (Dong et al., 2015), as shown in Fig. 2 (b); and (ii) silicon carbide, representative of elastic brittle ceramics, with $E_s$ = 410 GPa, $\sigma_f$ = 550 MPa, $v_s$= 0.14, and no strain hardening (resulting in $\varepsilon_f = \sigma_f/E_s$ = 0.0013), as shown in Fig. 2 (c).

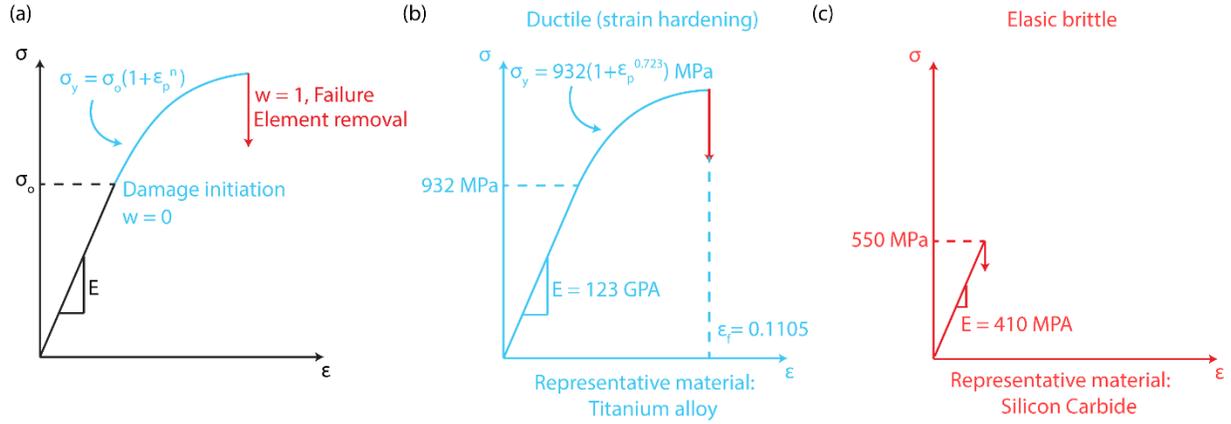

*Fig. 2. Representative stress-strain curves of (a) constitutive and damage model, (b) titanium alloy (Ti-6Al-4V), and (c) silicon carbide.*

## *2.4 Finite element methodology for extraction of the fracture toughness*

Mode-I fracture toughness of SENB specimens made of triangular, kagome, hexagonal lattices at $\bar{\rho}$ = 13%, 16% and 20%, and $\ell$ = 3 mm and 10 mm was investigated by simulating three point bending experiments (as outlined in ASTM E1820 (ASTM E1820, 2011)) by finite elements analysis. We used Abaqus/Explicit for all simulations, with mass scaling appropriately chosen to approach quasi-static response. To represent two-dimensional bars with rectangular cross sections in plane strain with one-dimensional beam elements, $E_s$ and $v_s$ of the base materials were replaced by modified Young's modulus, $E_s' = E_s/(1 - v_s^2)$ and Poisson's ratio, $v_s' = v_s/(1 - v_s)$ (Fleck and Qiu, 2007; Timoshenko and Woinowsky-Krigerm, 1959). Three spreaders (defined as analytic rigid surfaces in Abaqus) were used to apply three-point-bending boundary conditions without suffering local indentations; this practice had been used previously for fracture toughness testing of strut-based lattices (O'Masta et al., 2017; Quintana-alonso et al., 2010) and was also recommended by ASTM C393/C393M (ASTM C393/C393M, 2016) for flexural testing of sandwich beams. Frictionless contacts were assumed between the spreaders and the specimen, to further reduce localized deformation. One spreader was placed at the top midspan above the sample notch, while the other two spreaders were placed four sample widths apart at the bottom of the specimen, as shown in Fig. 3 (a).

Boundary conditions were applied as follows: (i) each bottom spreader could rotate around a fixed reference point, representing the center of a roller, and (ii) a sufficiently large load-line displacement was applied at the top spreader in the negative y-direction, until the final crack length is greater than the maximum crack capacity for fracture toughness calculations specified by ASTM E1820 (ASTM E1820, 2011), as shown in Fig. 3 (a). The damage model with maximum axial strain criterion (refer to Section 2.3) was used to describe the local fracture around the crack tip. For this criterion, accurate calculation of the axial stresses and strains in each beam is essential. For bending-dominated lattices, the maximum axial strain at the surface of each beam is significantly higher than the value at the neutral axis. As the default stress outputs for beam elements are at the neutral axis, an Abaqus user subroutine VUSFLD (vectorized user defined field) was created to extract beam surface stress and strain levels. As high frequency oscillations of the mechanical response could occur during a fracture event, we applied Butterworth real-time filtering with a cutoff frequency equals to the twice natural frequency of the specimen, extracted during quasi-static analyses. The natural frequency of each specimen was obtained by performing a linear perturbation analysis (see Appendix C).

Following the procedures outlined for fracture toughness calculations in ASTM E1820 (ASTM E1820, 2011), and summarized in Appendix D, the load-line displacement and the load-line reaction force were extracted at the top spreader and used to calculate the J-integral, $J$, consisting of both elastic and inelastic contributions to crack resistance. As the specimens have simple 2D prismatic architectures, crack extensions were tracked visually in post-processing. The fracture toughness, $K_J$, was then calculated as $K_J = \sqrt{EJ}$ where $E$ is the effective Young's modulus of the lattice material, simply extracted from the scaling laws of triangular, kagome, and hexagonal lattices (Fleck et al., 2010; Fleck and Qiu, 2007). The details of all fracture toughness calculation procedures are summarized in Appendix D. R-curves were constructed by plotting $K_J(i)$ against the change of crack length at instant (i), i.e., $\Delta a(i) = a(i) - a(1)$, and fitted with a two-term power law equation:

$$K_J = K_{JIc} + c\left(\frac{a(i)-a(1)}{a(1)}\right)^m \quad (1)$$

where $K_{JIC}$ is the initial fracture toughness, $c$ is the toughening coefficient, and $m$ is the toughening power.

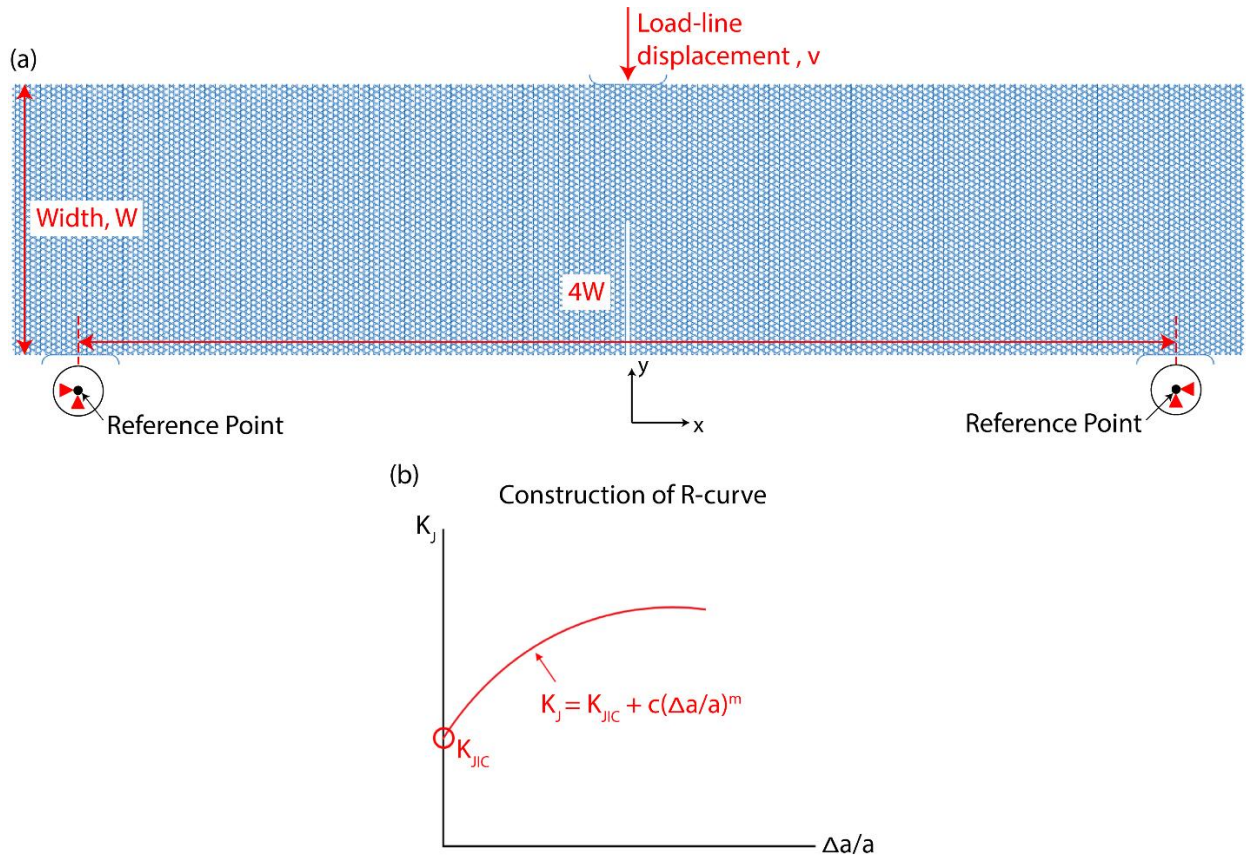

*Fig. 3. (a) Boundary conditions on a SENB specimen made of 2D lattices. Three spreaders are used: one at the top midspan and two at the bottom locations distance 4W from each other, to prevent highly localized deformation. Load-line displacement are applied at the top spreader while*

each of the bottom spreaders can rotate around a reference point, representing the center of a roller. (b) Construction of R-curve with fracture toughness, $K_J$ against normalized change of crack length, $\frac{\Delta a}{a}$. Two-term power law fit, $K_J = K_{JIC} + c(\frac{\Delta a}{a})^m$, where $K_{JIC}$ is the initial fracture toughness, $c$ is the toughening coefficient, and $m$ is the toughening power.

**3. Fracture toughness of 2D lattices**

*3.1 Mechanisms of deformation and damage and the size of the plastic region*

*(a) Silicon carbide specimens*

The load-line reaction force, $P$, versus load-line displacement, $v$, curves for the three cellular specimens made of elastic brittle silicon carbide are shown in Fig. 4 (a). Two different relative densities are explored for each topology. Filtered and unfiltered curves are identical until the initial fracture event, at which point the unfiltered $P$-$v$ curves start to display high frequency response. In general, deformations are elastic brittle, such that $P$ increases linearly with $v$ until initial fracture (denoted as point 1 in Fig. 4 (a)), followed immediately by catastrophic failure (significant loss of load-carrying capacity). The exception is the triangular specimen at $\bar{\rho} = 20\%$, which retains most of its load-carrying capacity after initial fracture and requires additional, *albeit* small, load to come to catastrophic failure. We hypothesize that this unusual behavior is likely due to crack deflection and crack branching occurring right after the initial fracture event, as shown in Fig. 4 (c): after the first bar, which is aligned with the principal stress direction, breaks, the crack hits a node and the next failure event occurs at side bars which are slightly less loaded, requiring additional work to fracture. Clearly, this behavior can be modified by node design, which is absent in the FEM simulations presented herein. The implication is that careful architectural design can potentially increase the fracture toughness, even in the case in which the base material is elastic brittle. In Fig. 4 (b), zoomed-in crack tip regions are drawn for each specimen to indicate initial fracture locations.

These initial fracture locations agree well with those identified using boundary layer analyses in (Fleck and Qiu, 2007).

*(b) Titanium alloy specimens*

The specimens made of elastoplastic titanium alloy (Ti-6Al-4V) initially deform in a linear elastic manner, followed by yielding, strain hardening and subsequently initial fracture; afterwards, the specimens experience gradual damage propagation (rather than abrupt catastrophic load drop as in elastic brittle base materials), up to final failure (defined as significant loss of load-carrying capacity). As for the case of the brittle materials, both the filtered and unfiltered P- v curves are identical until the initial fracture events.

Triangular lattices exhibit the most gradual transitions to final failures: after the initial fracture (denoted as point 1 in Fig. 5 (a)), $P$ still increases with v to a peak load followed by multiple stages of small load drops (denoted as points 2, 3, 4, and 5 in Fig. 5(a)). Each load drop is determined by fracture of a few bars, with the precise sequence indicated in Fig. 5(d). Kagome lattices display the second most gradual transition to final failure: after initial fracture (denoted as point 1 in Fig. 5 (b)), $P$ increases with $v$ up to a peak load, followed by a couple of load drops (denoted as points 2 and 3 in Fig. 5(b)); each drop corresponds to fracture of a number of bars, which is generally larger than for triangular lattices, as shown in Fig. 5(e). Finally, hexagonal lattices show the most sudden transition to final failure: after the initial fracture event (denoted as point 1 in Fig. 5 (c)), $P$ increases with $v$ up to a peak load, but is immediately thereafter followed by a catastrophic load drop (denoted as point 2 in Fig. 5 (c)). In addition, the corresponding crack extends and branches along two paths (both at a 45-degree angle to the horizontal) as shown in Fig. 5 (f). Interestingly, the hexagonal lattice is the only lattice that exhibits crack branching.

(a) 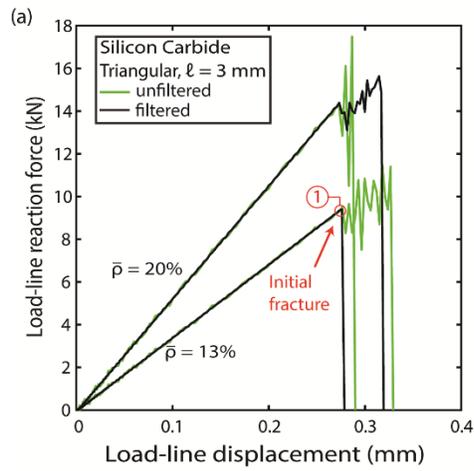 (b) 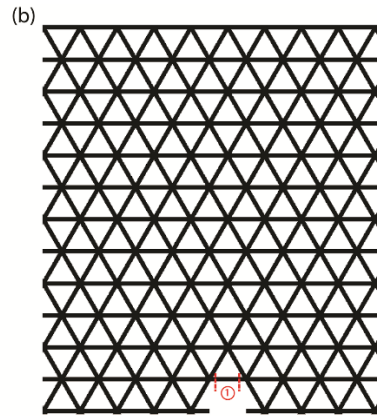

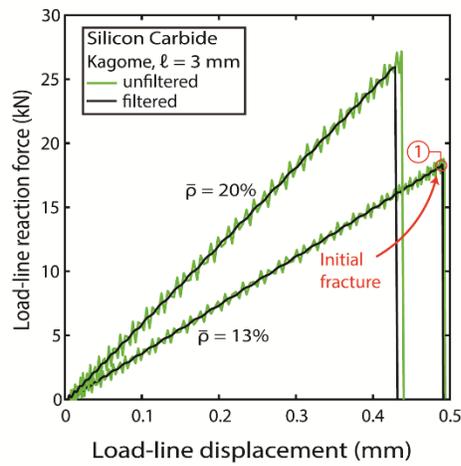 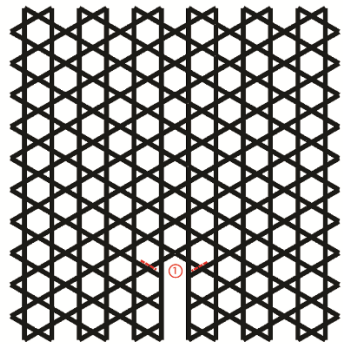

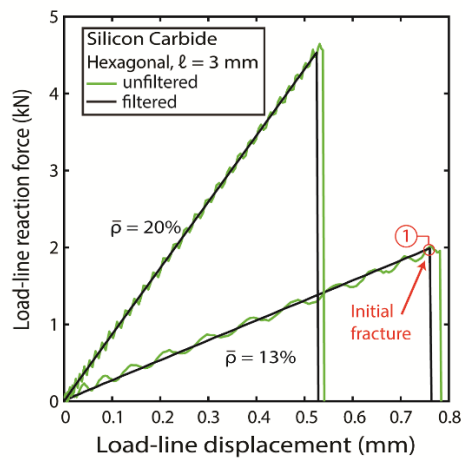 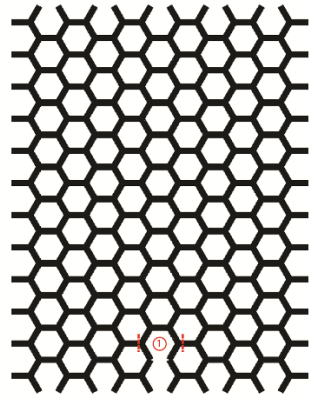

(c) 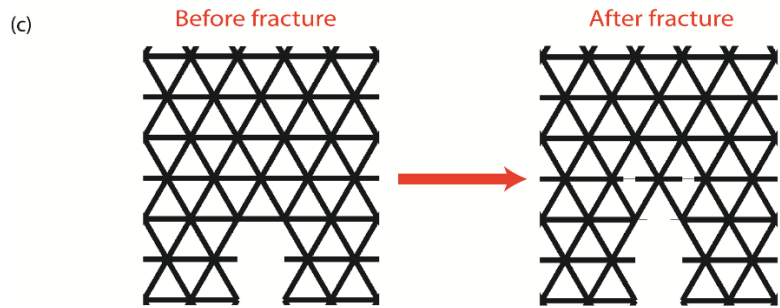

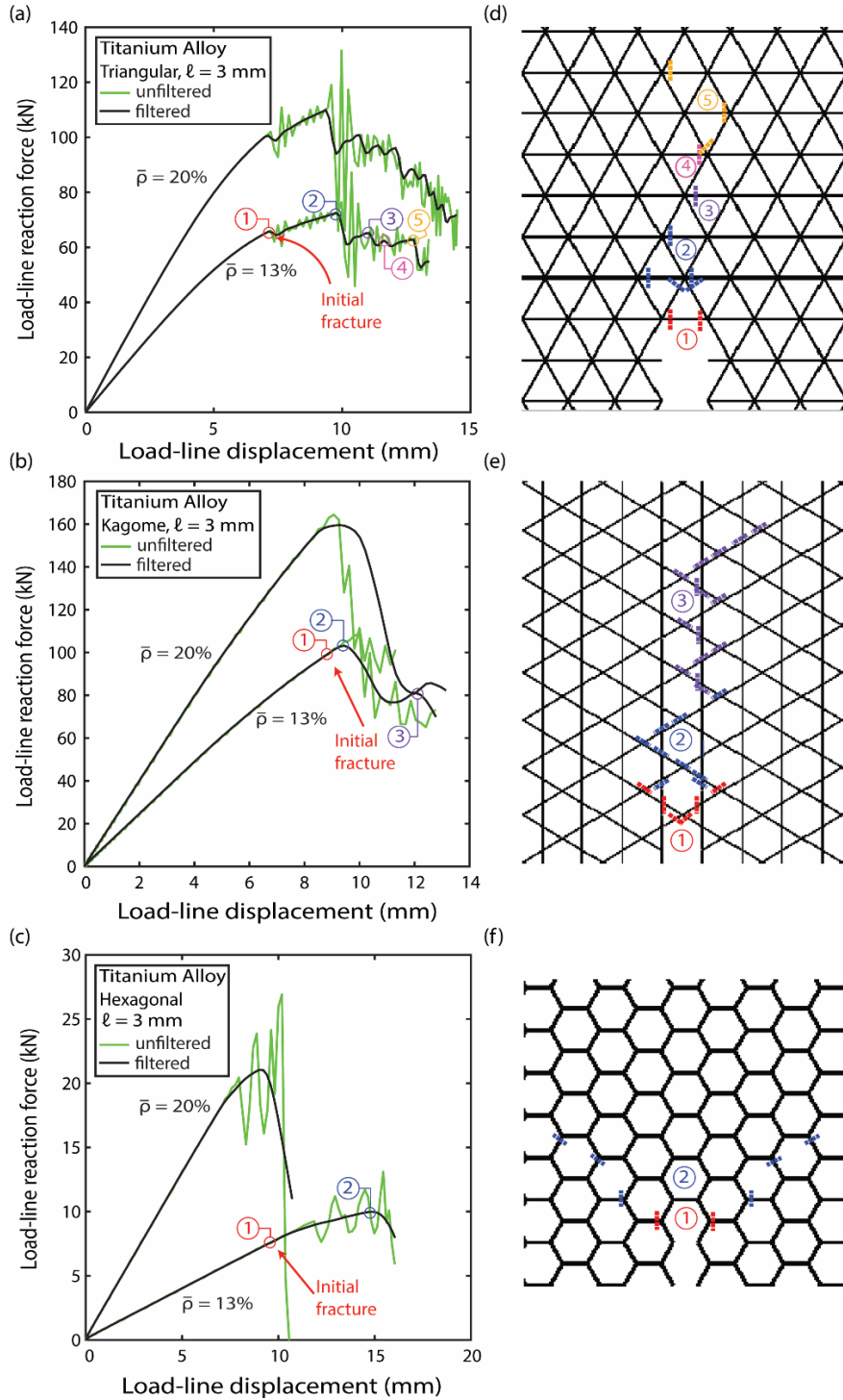

*Fig. 4. Deformation maps of topological specimens made of silicon carbide at $\bar{\rho}$ =13% and 20%: (a) load, P, versus load-line displacement, v, were plotted till overall kinetic energy exceeds 5% of overall internal energy of each specimen in each simulation. Encircled number one corresponds to the initial fracture followed immediately by catastrophic failures (significant load drop) and(b) the corresponding initial fracture locations are represented by red dashed line (c) Illustration of crack branching of the $\bar{\rho}$ =20% triangular lattice immediately after fracture. All lattices are not drawn to scale.*

*Fig. 5. Deformation maps of topological specimens made of Titanium Alloy (Ti-6Al-4V) at $\bar{\rho} = 13\%$ and 20%: load, P, versus load-line displacement, v, of (a) triangular specimens, (b) kagome lattices, and (c) hexagonal lattices were plotted till overall kinetic energy exceeds 5% of overall internal energy of each specimen in each simulation. Encircled number one corresponds to initial fracture and the subsequent encircled numbers correspond to each load drop. The fracture locations of each encircled number in (a), (b), and (c) are represented by corresponding colored dashed line in (d), (e), and (f) respectively; the lattices are not drawn to scale.*

These differences in crack propagation can be directly related to the size of the process zone ahead of the crack tip. The larger the process zone (i.e., the number of bars that are experiencing significant plastic deformation during crack propagation), the more gradual the loss of load carrying capacity in the sample, and the more pronounced the toughening response of the lattice. For the purpose of simple comparison between three lattices, we arbitrarily define the process zone as the rectangular region encompassing all the bars that attain the plastic strain limit, $\varepsilon_{plimit} = 0.01$. The process zone size right before the initial fracture event for the three sample topologies at two relative densities is plotted in Fig. 6. Regardless of the relative density, the triangular specimens have the largest fractional area (defined as the process zone area normalized by the sample size) of 2.4%, about twice as large as that for kagome specimens; by contrast, the hexagonal specimens have extremely localized failures, with negligible process zone size.

(a)
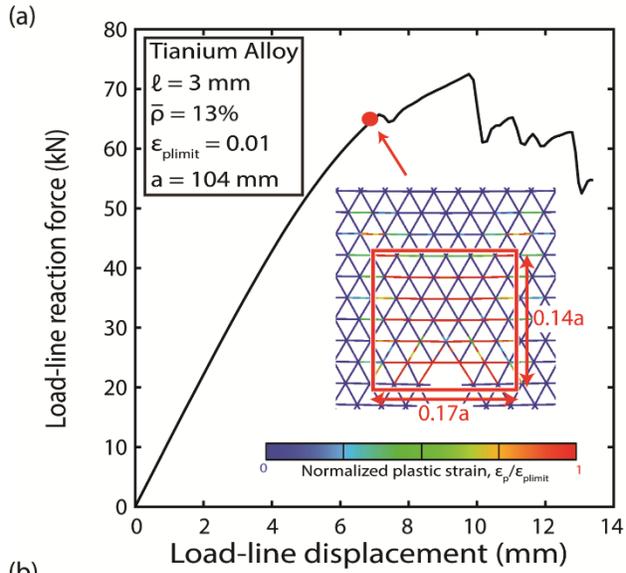
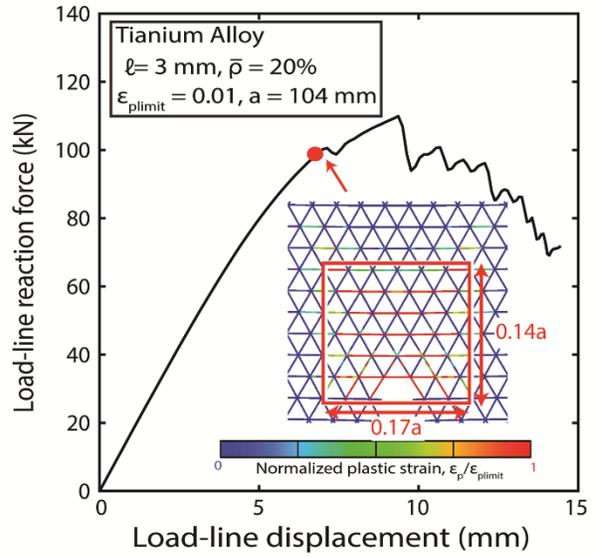

(b)
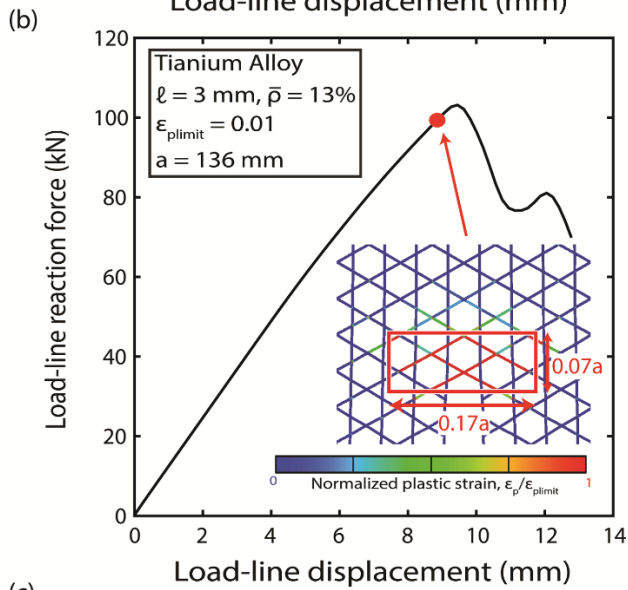
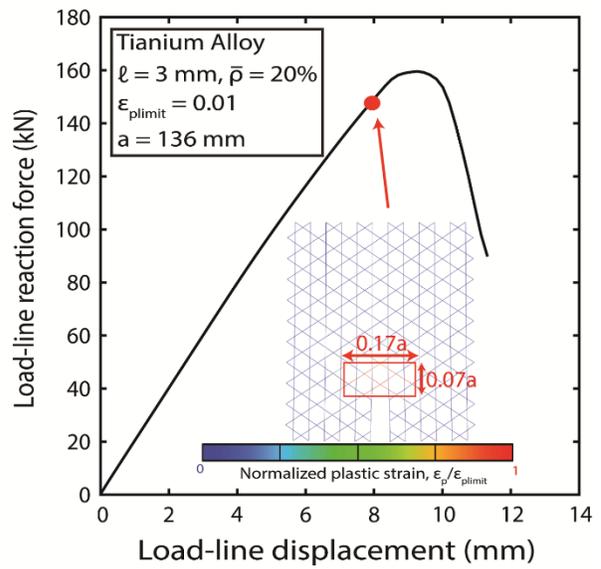

(c)
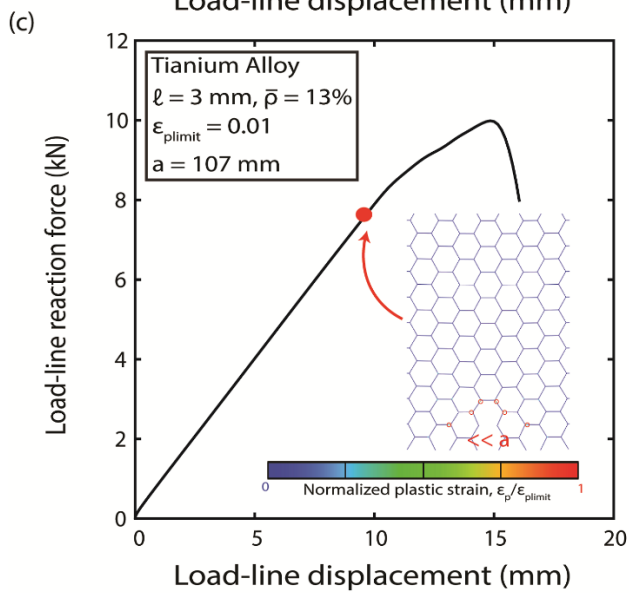
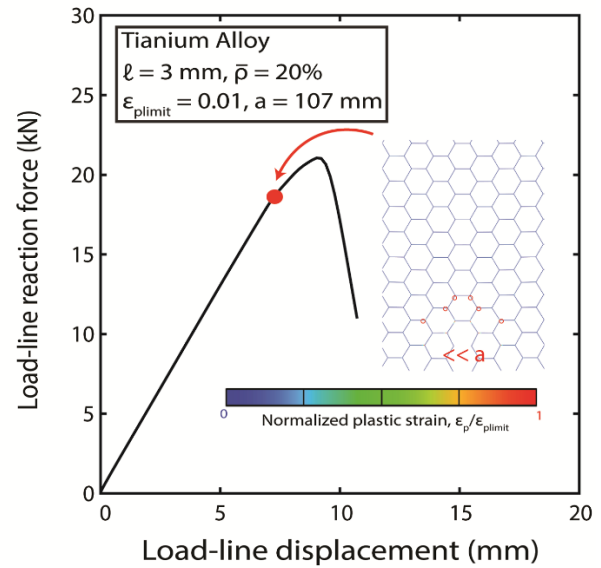

*Fig. 6. Contours of absolute normalized plastic strain, $\varepsilon_p/\varepsilon_{plimit}$, around the crack tip between (a) triangular lattices, (b) kagome lattices, and (c) hexagonal lattices made of titanium alloy (Ti-6Al-4V) with $\ell = 3$ mm at relative density, $\bar{\rho} = 20\%$ and $13\%$ right before the initial fracture. The corresponding location on the load-displacement curve of each contour is indicated by a red dot. Plastic strain limit, $\varepsilon_{plimit} = 0.01$ is used in the strain contours such that for all $\varepsilon_p/\varepsilon_{plimit} > 1$, the contour appears to be red. Red rectangles or circles are used to estimate the size of plastic regions. The inset lattices are not drawn to scale.*

### *3.2 Scaling relations for fracture toughness and R-curve*

#### *(a) Silicon carbide specimens*

The fracture toughness of elastic brittle specimens was extracted using the J-integral method, only considering the elastic $J$ component, $J_{el}$, (see Appendix D for details). The fracture toughness, $K_{IC}$, scales with relative density, $\bar{\rho}$ as $\bar{\rho}^1$, $\bar{\rho}^{0.5}$, and $\bar{\rho}^2$ in triangular, kagome, and hexagonal specimens, respectively, as shown in Fig. 7 (a). In addition, we found that the fracture toughness is largest for the Kagome specimens and lowest for the hexagonal specimens. Furthermore, we verified that $K_{IC}$ depends linearly on the square root of the strut length, $\sqrt{\ell}$ (Maiti et al., 1984), as shown in Fig. 7 (b). All these findings are perfectly consistent with the results obtained using boundary value analysis in (Fleck and Qiu, 2007), and thus validate our numerical approach for the calculation of the fracture toughness in cellular materials specimens.

#### *(b) Titanium alloy specimens*

The initial fracture toughness, $K_{JIC}$, toughening coefficient, $C$, and toughening power, $m$, of the titanium alloy (Ti-6Al-4V) lattice specimens were extracted following the procedures discussed in section 2.4.

The initial fracture toughness reveals some important results: (i) while the scaling relationships for $K_{JIC}$ with $\bar{\rho}$ remain the same as for the elastic brittle case, the actual values of $K_{JIC}$ are ~5 times larger, as shown in Fig. 8 (a); (ii) the difference between the performance of the triangular and kagome lattices is reduced relative to the brittle base material case, while the hexagonal lattice remains the least efficient design. The implication is that high initial fracture toughness can be achieved at low relative density by designing a triangular or kagome lattice made of a ductile base material. In addition, we show that even in the plastic case, $K_{JIC}$ still linearly depends on $\sqrt{\ell}$, as shown in Fig. 8 (b). The superior efficiency of the triangular lattices can be attributed to the extent of the process zone ahead of the crack tip (Fig. 6), which more than compensates the elastic crack blunting that makes kagome designs preferred for brittle materials (Fleck and Qiu, 2007).

Similar scaling relationships can be extracted for the toughening coefficient, $C$, as shown in Fig. 9 (a). Remarkably, the scaling exponents are the same as for the initial fracture toughness. Triangular lattices have the largest toughening coefficients, followed by kagome and hexagonal lattices. By contrast, the toughening exponent, $m$, does not substantially depend on the relative density or lattice topology, maintaining a value ~0.52-0.58. These features are evident in the R-curves displayed in Fig. 9 (b). Notice the substantial difference between triangular and kagome lattices: while their initial fracture toughness is very similar, triangular lattices toughen much more strongly during crack propagation. Again, this can be attributed to the larger process zone dimension in triangular specimens. Consistently with their very isolated yielding events, hexagonal lattices display much lower fracture toughness than the other lattice classes, especially at low relative density. We should emphasize that our choice to define local strut fracture when a critical value of the strain is reached at the outer surface of the strut potentially penalizes bending-dominated lattices made of ductile base materials, in which struts can still carry load after surface cracking has initiated. As hexagonal lattices are bending dominated, while triangular lattices are stretching dominated (with kagome lattices in between), this factor might contribute to the large performance difference among the three lattice topologies examined here. While in practice hexagonal and kagome lattices might perform a bit better than predicted here, we do not expect this contribution (if present) to affect any of the conclusions of this work.

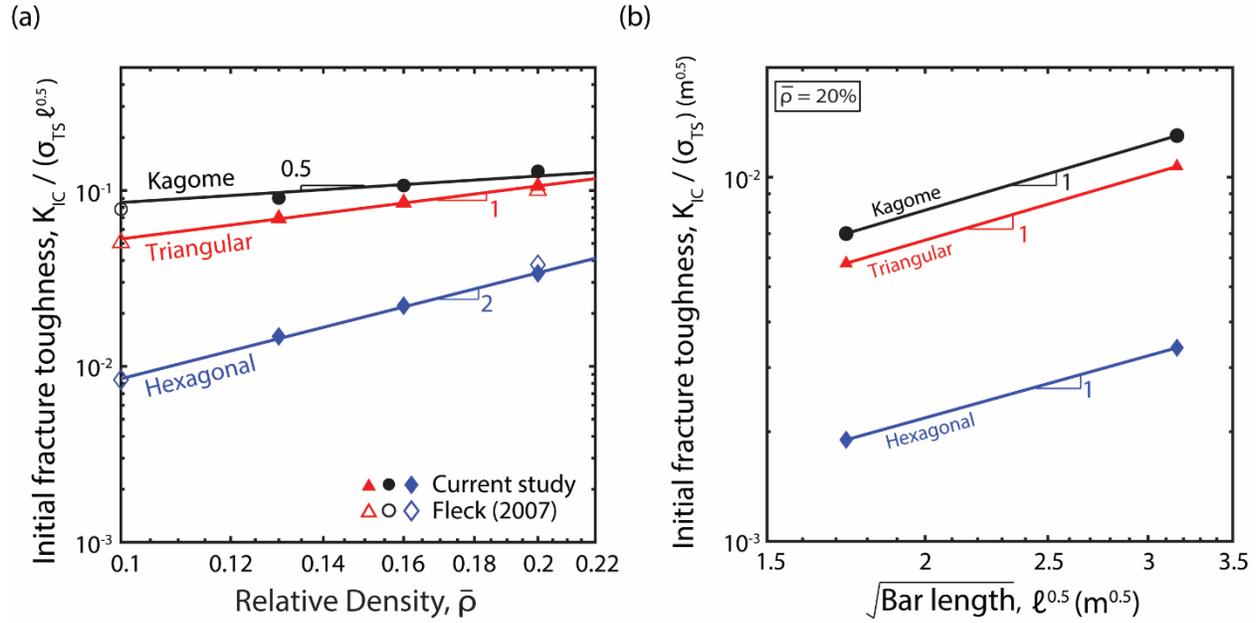

*Fig. 7. Scaling relationships for initial fracture toughness in triangular, kagome, hexagonal specimens made in silicon carbides with (a) the relative density $\bar{\rho}$ and (b) the square root of bar length, $\sqrt{\ell}$.*

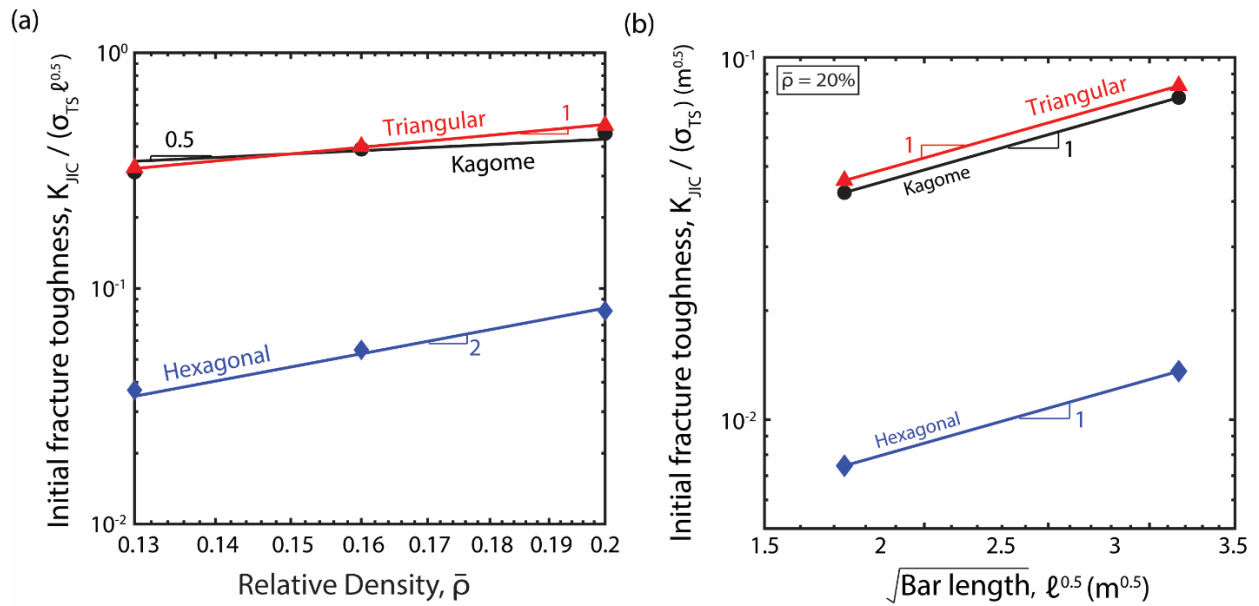

*Fig. 8. Scaling relationships for initial fracture toughness in triangular, kagome, hexagonal specimens made in titanium alloys (Ti-6Al-4V) with (a) the relative density $\bar{\rho}$ and (b) the square root of bar length, $\sqrt{\ell}$.*

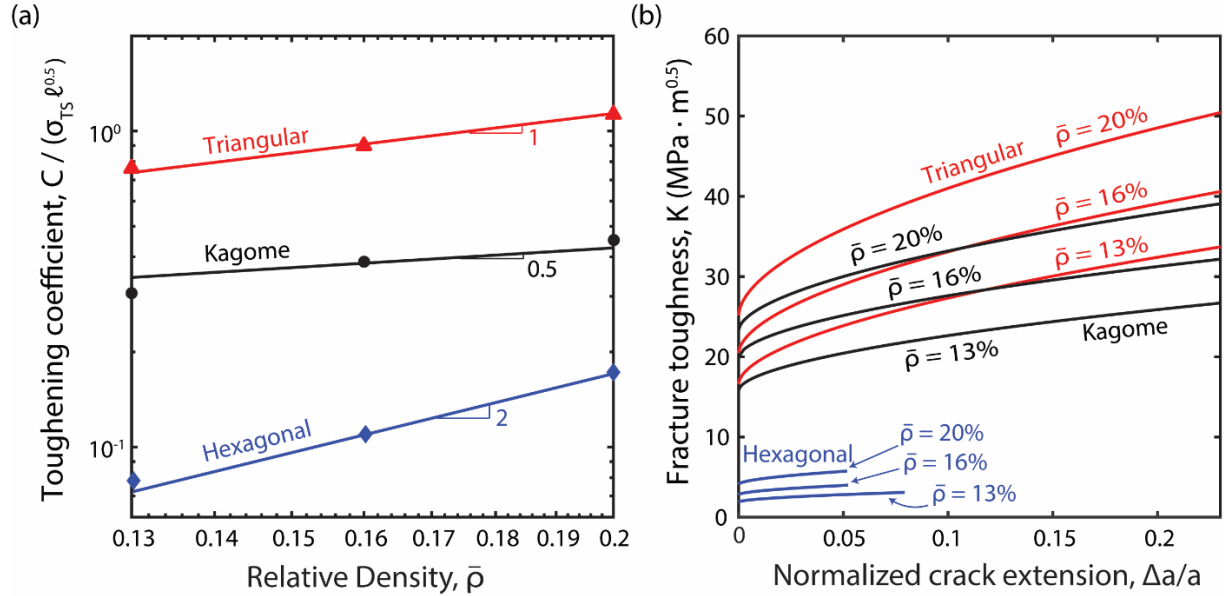

*Fig. 9. (a) scaling relationships for toughening coefficients, C, and (b) R-curves of triangular, kagome, and hexagonal specimens made in titanium alloy (Ti-6Al-4V).*

It is instructive to compare the fracture toughness of lattices with the universe of existing materials. We plot both initial fracture toughness, $K_{JIC}$, and fracture toughness at final failure, $K_{JFC}$, of triangular, kagome, and hexagonal lattices of $\ell = 3$ mm at $\bar{\rho} = 13\%$, $16\%$, and $20\%$ (corresponding to density of 585-900 kg/m$^3$) made in titanium alloy ($E_s = 123$ GPa, $v_s = 0.3$, $\sigma_{ys} = 932$ GPa, $n = 0.7237$, $\varepsilon_f = 0.1105$) as shown in Fig. 10. Both triangular and kagome lattices at this length scale outperform the majority of natural materials in terms of $K_{JIC}$. When considering $K_{JFC}$, triangular lattices approach the fracture toughness of bulk titanium alloy. In contrast, hexagonal lattices are comparable to other natural materials (e.g, paper) in terms of both $K_{JIC}$ and $K_{JFC}$. The conclusion is that topology manipulations can significantly alter both $K_{JIC}$ and $K_{JFC}$. However, this conclusion is here only proved for 2D lattices, and remains unexplored for 3D topologies, either strut-based or shell-based.

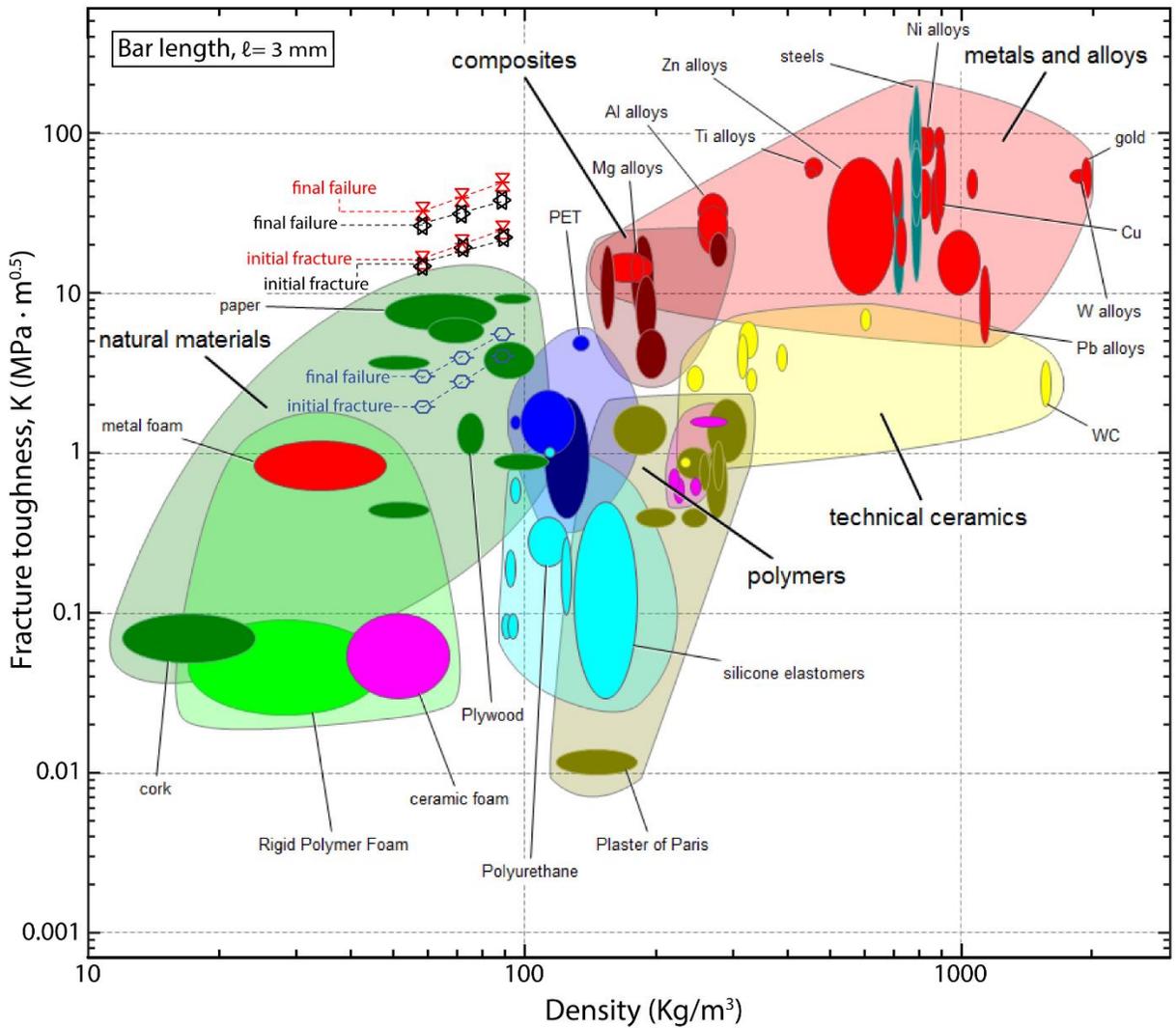

*Fig. 10. Material space charts for fracture toughness versus density. Three lattices are shown by their representative unit cell topologies: triangular (red), kagome (black), and hexagonal (blue). The prediction is based on titanium alloy material properties with the bar length, $\ell = 3$ mm.*

## 4. Discussion

In the case of elastic brittle base materials, the fracture toughness is governed by linear elastic fracture mechanics (LEFM) and in general depends linearly on the tensile strength of the base material. The fracture toughness of lattices predicted from the proposed numerical approach is in good agreement with solutions derived using boundary layer analyses for elastic brittle lattices as shown in Fig. 7 (Fleck and Qiu, 2007). The 24% discrepancy in the pre-factor, $D$ (see Table 1) for kagome lattices between this work and Fleck and Qiu, 2007 is attributed to different density ranges over which the parameter was fitted ($\bar{\rho}$ = 13% - 20% in the current analyses VS $\bar{\rho}$ = 0.3% - 20% in (Fleck and Qiu, 2007), and the larger sensitivity of the parameter D over the density range in kagome lattices, due to the elastic blunting zone around the crack tip (Fleck and Qiu, 2007).

In the case of elastoplastic constituent materials, the fracture toughness is still governed by LEFM if the plastic region size in front of crack tip is negligible compared to the crack length, i.e., small-scale yielding (SSY); conversely, elasto-plastic fracture mechanics (EPFM) governs if the plastic region size in front of the crack tip grows comparable to the crack length. The simple criterion for the SSY conditions to be satisfied in SENB specimens can be derived by combining the ASTM specification that $a, B > 2.5(K_I/\sigma_{ys})^2$ and the expression for the plane strain plastic zone size in a perfectly plastic solid, i.e., $r_p = \frac{1}{3\pi}(K_I/\sigma_{ys})^2$, resulting in the condition $a, B > 25r_p$. The plastic region sizes right before initial fracture in three lattices (Fig. 6) reveal that only the hexagonal lattice satisfies the SSY criterion. Hence the initial fracture toughness, $K_{JIC}$, of hexagonal lattices derived from current numerical approach should be comparable to the results obtained using K-field boundary layer analyses under small-scale yielding conditions in (Tankasala et al., 2015). For this reason, the $D$ pre-factors for the hexagonal lattices obtained with the two approaches agree well for hexagonal lattices but deviate for triangular and kagome lattices (Table 1). These findings further validate the current numerical approach for elastoplastic materials.

Finally, we emphasize that the numerical approach proposed herein can be easily extended to 3D cellular materials (strut or shell-based) with small modifications on the sample design (number of unit cells, number of elements, degree of isotropy) and boundary conditions. The only essential assumption is that the crack tip must be within the empty space of a unit cell, in order to avoid stress singularity and hence mesh-dependent damage in the finite element analysis. In the case of shell-based cellular materials, care must be taken to ensure the shell surfaces are sufficiently smooth to avoid sharp corners. We recognize two limitations of the proposed approach: (i) SENB specimens can only be used to predict mode-I fracture toughness, and (ii) progressive failure of individual struts can only be properly addressed by using continuum elements (not beam element as done in the current study), with significant increase in computational expense.

## 5. Conclusions

We proposed a versatile numerical approach for fracture toughness and R-curve modeling of both brittle and ductile cellular materials, combining J-integral calculations in SENB specimens and a local maximum strain damage model. The proposed model does not impose small-scale yielding restrictions, and hence does not require exceeding large numbers of unit cells for highly deformable constituent materials. This approach was applied in finite element modeling to investigate fracture toughness (R-curves) of three 2D isotropic lattices: triangular, kagome, and hexagonal lattices, made of both silicon carbide (elastic brittle) and titanium alloy (elastoplastic) constituent materials. First, specimen design studies were systematically conducted in order to extract the required number of unit cells, number of elements per bar and degree of isotropy, in order to ensure that accurate stress/strain fields are captured around the crack tip. Second, we validated the current numerical approach by comparing the predicted initial fracture toughness in both elastic brittle and small-scale yielding elasto-plastic cases with previous works (Fleck and Qiu, 2007; Tankasala et al., 2015). Finally, we extracted four key results: (i) toughening via topological designs of the lattice architecture seems possible even in elastic brittle lattices; (ii) the power-law scaling relations for initial fracture toughness and toughening coefficients with relative densities only depend on the topology of the lattice, and not the mechanical response of the base material; (iii) triangular lattices outperform both kagome and hexagonal lattices in terms of toughening (R-curve) in the range of $\bar{\rho} = 0.13 – 0.2$, and (iv) careful design of topology to exploit toughening mechanisms in cellular materials can significantly increase the steady-state fracture toughness that possibly exceed the initial fracture toughness or even the steady-state fracture toughness of the constituent material itself (especially true in the case of brittle base materials). We emphasize that the strength of the proposed approach lies in its versatility, allowing analysis of both 2D and 3D lattice materials with virtually any topology (both truss and shell-based) and any constituent material behavior, with the only restriction that the crack tip must be contained inside the unit cell. This topological condition could be relaxed with further development. These features make the proposed approach ideally suitable for the toughness calculation of shell-based lattice materials.

**Appendix A. Extraction of the minimum number of unit cells for accurate toughness modeling**

Linear elastic, small-strain finite element simulations were performed to calculate the minimum number of unit cells along the y-direction, $n_y$, of SENB specimens in order to secure accurate toughness measurements. The overall specimen dimensions follow the ASTM E1820 standard (ASTM E1820, 2011), with aspect ratio of $S = 4W$, $W = 2B$ and $a/W = 0.5$, as shown in Fig. A1 (a). Displacement-controlled boundary conditions were applied as follows (see Fig. A1(a)): (i) A unit displacement of 1 mm was applied along the negative y-direction on the top surface node at the mid-span of the specimen; (ii) Two pin-jointed supports were applied at bottom nodes, distanced $W/4$ away from each end of the specimen, to prevent displacement in the y-direction. (iii) A fixed boundary condition was applied at one of the pin-jointed supports, to prevent rigid body motion in the x-direction. Each bar is represented by a single Euler-Bernoulli beam element (element B23 in Abaqus) with cubic interpolation; these elements can capture the deformation of a slender bar subjected to arbitrary bar end moments and forces in a small strain analysis. We varied the number of unit cells in the y-direction from $n_y = 4$ to $n_y = 45$, while keeping the unit cell size at $\ell = 3$ mm and relative density at $\bar{\rho} = 0.1$. The pre-factor, $D$, for each topology was then calculated as $D = K_{IC}/(\sigma_{TS}\bar{\rho}\sqrt{\ell})$. As these are small-displacement linear elastic analyses, $K_{IC}/\sigma_{TS}$ can be replaced by $K_{IC}/\sigma_{max}$, where $K_I$ is the imposed mode-I stress intensity factor and $\sigma_{max}$ is the maximum stress around the crack tip under the imposed displacement. Note that the constituent material properties are immaterial and would not change the results in these analyses. The normalized axial stress contours of triangular specimens with $n_y = 8$ and $n_y = 40$ were plotted in Fig. A1 (b). Maximum axial stress occurred immediately in front of the crack tip; the same conclusion can be drawn for hexagonal and kagome specimen (contours not shown here). The pre-factors $D$ for the three topologies were plotted against $n_y$, and the minimum $n_y$ for sufficient accuracy is then determined as the number of unit cells at which the pre-factor is within 5% of the final converged value. The chosen $n_y$ for hexagonal, kagome, and triangular lattices are 40, 45, and 40 respectively, as shown in red circles in Fig. A1 (c)

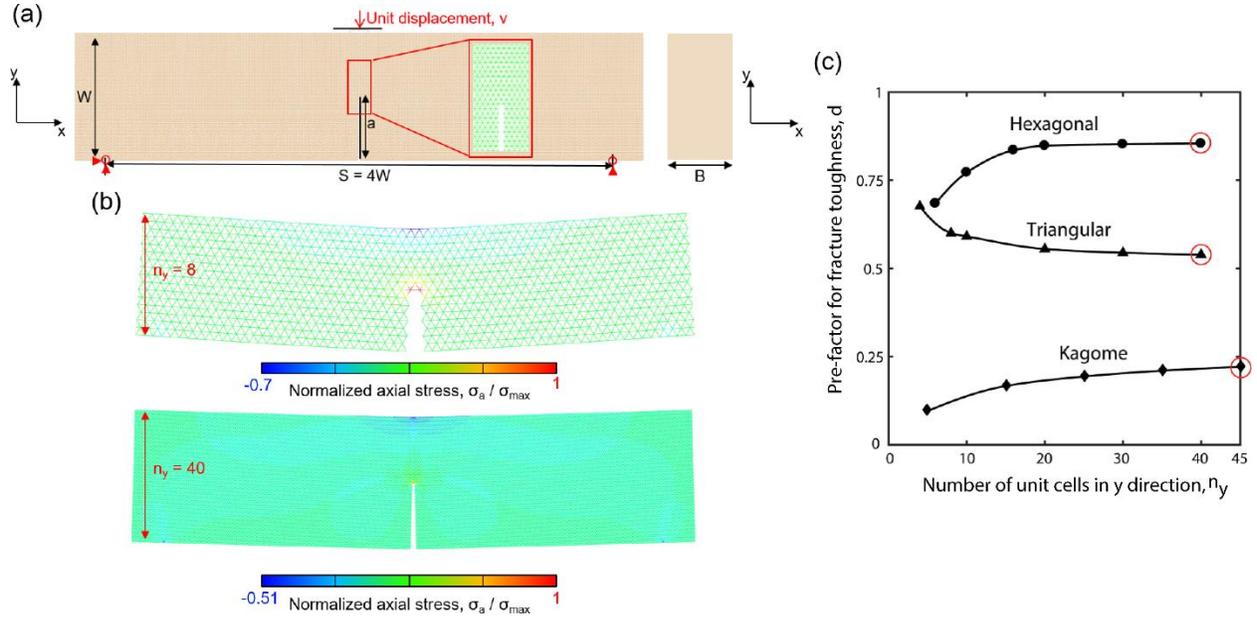

*Fig. A1. Specimen design: number of unit cells. (a) Specimen aspect ratio. (b) Normalized axial stress contour of triangular specimens at number of unit cells, $n_y$ = 8 and $n_y$ = 40. (c) Convergence of the pre-factor, D, with number of unit cells, $n_y$. The chosen $n_y$ for the simulations are circled in black.*

**Appendix B. Mesh sensitivity analysis**

Finite-strain elasto-plastic finite element simulations were performed to determine the number of elements per bar, $n_e$, required to accurately capture the initial fracture response of SENB specimens. The minimum number of unit cells in the y-direction, $n_y$, determined for each topology in Appendix 1 was used in all simulations, while keeping $\bar{\rho}$ = 13% and $\ell$ = 3 mm. The boundary conditions described in section 2.4 were applied. Each bar, representing a Mindlin-Reissner isotropic plate (Mindlin, 1951; Reissner, 1945), was modeled with first-order Timoshenko beam elements (element B21 in Abaqus) for large strain and rotation analyses. To increase computational efficiency, only a rectangular region spanning approximately 10 x 13 unit cells in each in-plane direction around the crack tip has refined mesh, varying from $n_e$ = 4 to $n_e$ = 50 as shown in Fig. B1 (a) and (b). We ensured that this rectangular region encompasses the maximum crack length extension specified in ASTM E1820 (ASTM E1820, 2011). Outside of the rectangular region, each bar is modeled with one Timoshenko beam element. Such coarse meshing far away from the crack tip would only affect local deformation and simulated results would give the fracture response of an ideally scaled-up specimens. Titanium alloy (Ti-6Al-4v) is chosen as representative material, with plane-strain-modified Young's modulus, $E_s' = 135$ GPa and modified Poisson's ratio, $v_s' = 0.43$; the plastic behavior is modeled as $\sigma = \sigma_{ys}(1 + \varepsilon_p{}^n)$, where $\sigma_{ys} = 932$ GPa is the initial yield strength, $n = 0.7237$ is the strain hardening power, and $\varepsilon_p$ is the plastic strain; a fracture strain $\varepsilon_f = 0.1105$ is assumed. The load-line displacement at which the maximum axial strain around the crack tip reached $\varepsilon_f$ was plotted against $n_e$. The minimum $n_e$ was then determined as the number of elements per bar at which the load-line displacement is within 10% of the final converged value. The chosen minimum $n_e$ for hexagonal, kagome, and triangular lattices are 30, 20, and 4 respectively, as shown in red circles in Fig. B1 (c).

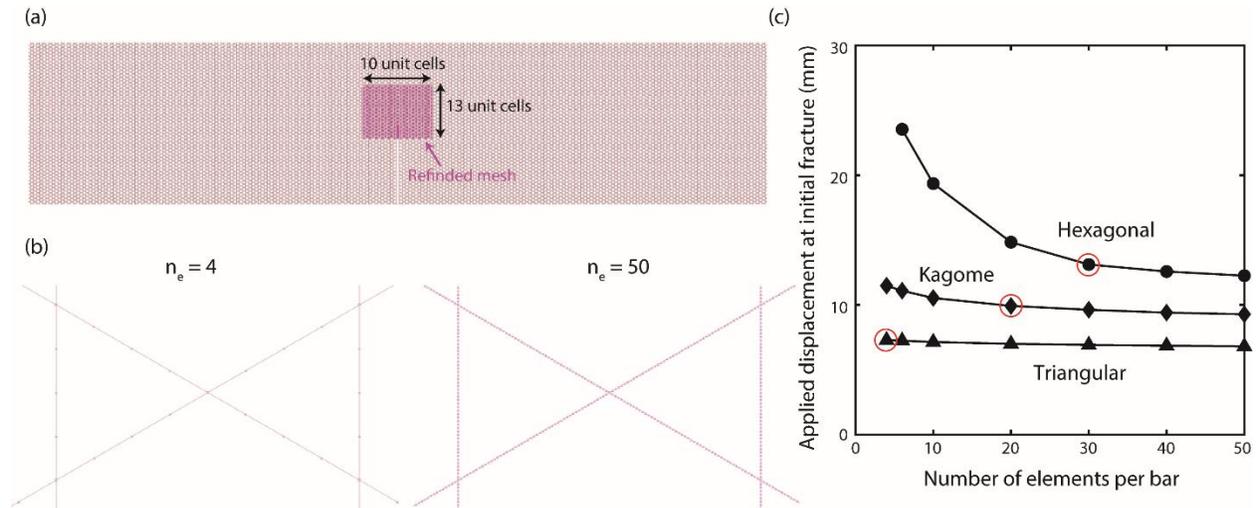

*Fig. B1. Specimen design: number of elements per bar. (a) Demonstration of refined mesh region around the crack tip. (b) Magnified views of mesh seeding around the crack tip with number of elements per bar, $n_e = 4$ and $n_e = 50$. (c) Convergence of the load-line displacement, at which the maximum axial strain around the crack tip reached $\varepsilon_f$, with number of elements per bar, $n_e$. The chosen $n_e$ for simulations are circled in red.*

## Appendix C. FE linear perturbation analysis to calculate the natural frequencies of 2D lattice materials

Linear perturbation finite element simulations were performed to calculate the natural frequencies, $\omega_n$, of all SENB specimens. Similar boundary conditions as defined in appendix A were applied, but without any loadings, as shown in Fig. C1. The minimum number of unit cells, $n_y$, and required number of elements per bar, $n_e$, derived from appendix 1 and 2 were used. Titanium alloy (Ti-6Al-4V) was used as constituent material, with modified Young's modulus, $E_s' = 135$ GPa and Poisson's ratio, $v_s' = 0.43$, and density $\rho_s = 4{,}500$ kg/m$^3$. The ten smallest eigen-frequencies with corresponding eigen-modes were first extracted in each simulation. The smallest eigen-frequency with a physical eigen-mode was then taken as the natural frequency. Natural frequencies at each relative density and topology are reported in Table C1. For lattice specimens made of silicon carbide and/or with mass scaling (artificial increase of density in Abaqus to speed up simulation), natural frequencies were inferred proportionally, using $\omega_n = (E_s'/\rho_s)^{0.5}$.

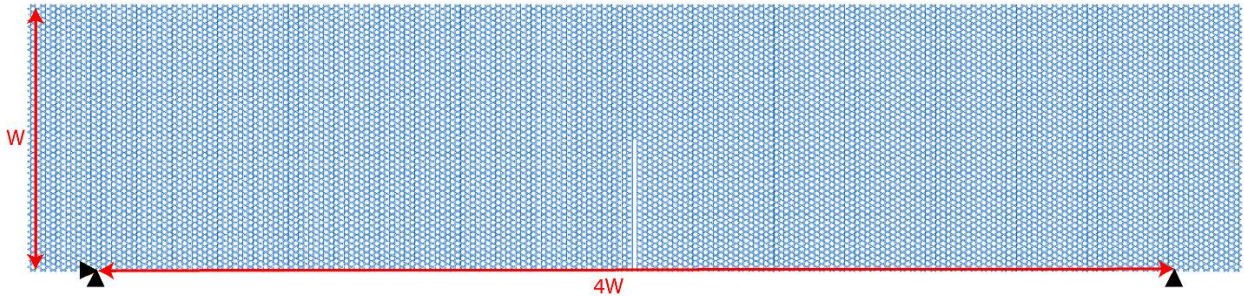

*Fig. C1. Illustration of boundary conditions applied in linear perturbation analyses: (i) two pin-jointed supports at 4W distance apart to prevent translate in the vertical direction and (ii) a fixed boundary condition applied at the left pin-jointed support to prevent rigid body motion.*

| Titanium Alloy Ti-6L-4V | Natural frequency, $\omega_n$ (cycles/s) | | |
| --- | --- | --- | --- |
| Topology | $\bar{\rho} = 13\%$ | $\bar{\rho} = 16\%$ | $\bar{\rho} = 20\%$ |
| Triangular | 202 | 202 | 202 |
| Kagome | 140 | 140 | 140 |
| Hexagonal | 51 | 61 | 74 |

*Table C1. Natural frequencies of titanium alloy (Ti-6AL-4V) lattice specimens made of triangular, kagome, and hexagonal topologies, at $\bar{\rho}$ = 0.13, 0.16, and 0.2. Base material properties of modified Young's modulus, $E_s'$ = 135 GPa and Poisson's ratio, $v_s'$ = 0.43, and density $\rho_s$ = 4,500 kg/m³ were used.*

**Appendix D. Fracture toughness calculation procedures**

Fracture toughness calculation procedures are taken from ASTM E1820 (ASTM E1820, 2011) and summarized below with minor modifications. Fracture toughness is first quantified in terms of J-integral, $J$. $J$ is composed of an elastic part, $J_{el}$ and a plastic part, $J_{pl}$:

$$J = J_{el} + J_{pl} \tag{D.1}$$

These components are calculated at instant (i), corresponding to each crack extension (load drop), as $J_{el}$ (i) and $J_{pl}$ (i) respectively. $J_{el}$ (i), is related to mode-I stress intensity factor, $K_I$ as follows:

$$J_{el}(i) = \frac{(K_I(i))^2}{E} \tag{D.2}$$

where E is the effective Young's modulus of the lattice specimens. $K_I$ (i) at each instant of crack extension is related to the load-line reaction force, P (i) as shown in Fig. D1 (a):

$$K_I(i) = \left[\frac{4P(i)}{B\sqrt{W}}\right] \cdot f\left(\frac{a(i)}{W}\right) \tag{D.3}$$

where *a (i)* is the crack length at the instant *(i)* and the calibration factor $f(a(i)/W)$ is calculated as,

$$f\left(\frac{a\,(i)}{W}\right) = 3\sqrt{\frac{a\,(i)}{W}} \cdot \frac{1.99 - \frac{a\,(i)}{W} \cdot \left(1 - \frac{a\,(i)}{W}\right) \cdot \left(2.15 - 3.93\frac{a\,(i)}{W} + 2.7\left(\frac{a\,(i)}{W}\right)^2\right)}{2\left(1 + 2\frac{a\,(i)}{W}\right) \cdot \left(1 - \frac{a\,(i)}{W}\right)^{1.5}} \tag{D.4}$$

As the specimens under investigation are 2D lattices, crack extensions can be visually determined in post-processing. $J_{pl}\,(i)$ is related to the area under the load-line force / displacement curve, $A_{pl}\,(i)$, at instant *(i)* as follows (see Fig. D1):

$$J_{pl}\,(i) = \frac{2A_{pl}\,(i)}{B \cdot (W - a\,(i))} \tag{D.5}$$

The fracture toughness at crack extension instant *(i)*, $K_J\,(i)$, is then calculated as $K_J\,(i) = (E \cdot J\,(i))^{0.5}$.

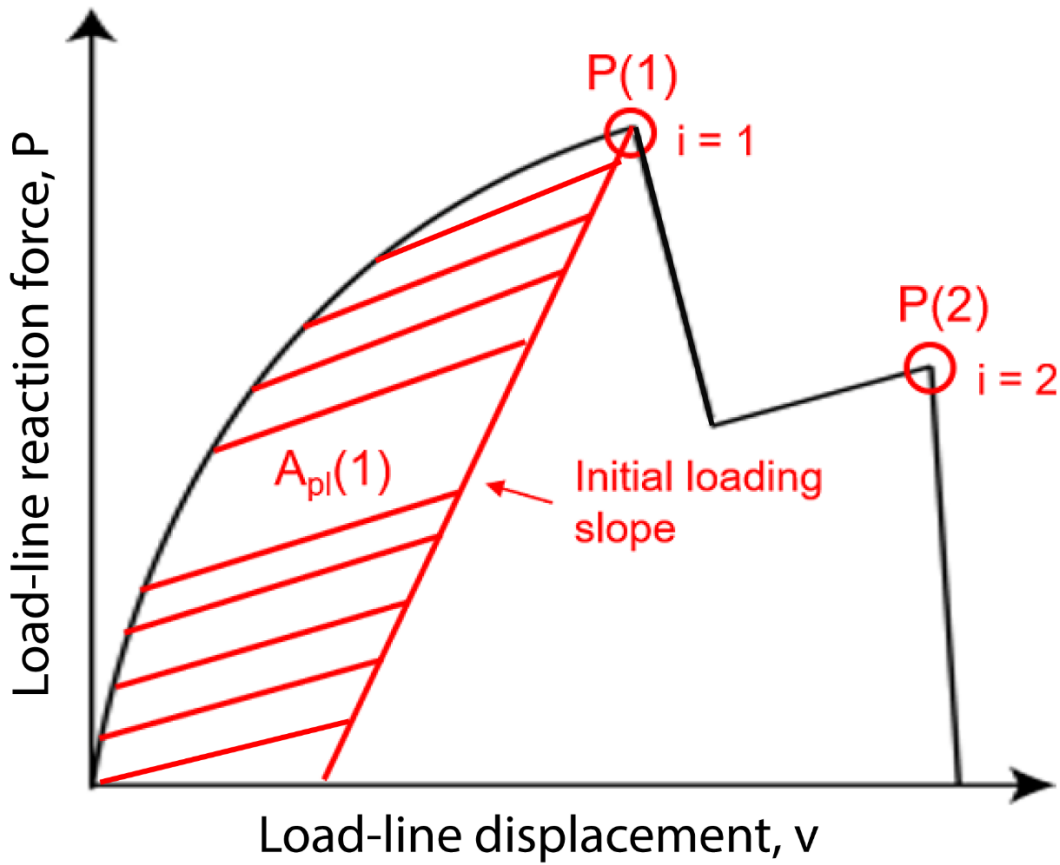

*Fig. D1. Illustration of load-line reaction force, P, versus load-line displacement, v, at crack extension instant i = 1 and 2 with the area under the curve, $A_{pl}$, at crack extension instant i = 1 for J-integral calculations.*

**Acknowledgments**

This research was supported by a NASA Early Stage Innovation (ESI) program, award # 80NSSC18K0259. We gratefully acknowledge the technical support and resources in high performance computing provided by the Research Cyber Infrastructure Center (RCIC) at UC Irvine. The ABAQUS Finite Element Analysis software is licensed from Dassault Systemes SIMULIA, as part of a Strategic Academic Customer Program between UC Irvine and SIMULIA.